\documentclass[12pt]{iopart}

\usepackage[dvips]{graphicx}
\usepackage{dcolumn}
\usepackage{bm}

\newcommand{\bdm}{\begin{displaymath}}
\newcommand{\edm}{\end{displaymath}}
\newcommand{\benl}{\begin{equation}}
\newcommand{\be}[1]{\begin{equation}\label{#1}}
\newcommand{\ee}{\end{equation}}
\newcommand{\bea}{\begin{eqnarray}}
\newcommand{\eea}{\end{eqnarray}}
\newcommand{\ba}{\begin{array}}
\newcommand{\ea}{\end{array}}

\newcommand{\qv}{\boldsymbol{\mathrm q}}

\newcommand{\kv}{\boldsymbol{\mathrm k}_\parallel}

\newcommand{\oH}{\widehat H}

\newcommand{\ket}[1]{|#1\rangle}
\newcommand{\bra}[1]{\langle#1|}
\newcommand{\scal}[2]{\langle #1|#2\rangle}
\newcommand{\media}[3]{\langle#1|#2|#3\rangle}

\newcommand{\de}{\mathrm{d}}
\newcommand{\deriv}[2]{\frac{\de #1}{\de #2}}

\newcommand{\bc}{\begin{center}}
\newcommand{\ec}{\end{center}}

\newcommand{\sch}{Schr\"odinger}

\begin{document}

\title{High-field transport in semiconductor superlattices for interacting 
Wannier-Stark  levels}

\author{Angelo Guida}
\address{Dipartimento di Fisica Universit\`a 
del Salento, Via Arnesano 73100 Lecce, Italy}

\author{Lino Reggiani}
\address{
Dipartimento di Ingegneria dell' Innovazione
and CNISM, Universit\'a del Salento, Via Arnesano s/n, 73100 Lecce, 
Italy}

\author{Marcello Rosini}
\address{
Dipartimento di Fisica Universit\`a di Modena e Reggio
Emilia and S$^3$ national research centre of CNR-INFM,
Via Campi 213/A, 4100 Modena, Italy}
\ead{marcello.rosini@unimore.it}

\date{\today}

\begin{abstract}
  We develop   a    microscopic  theory of  electron     transport  in
  superlattices within  the Wannier-Stark  approach  by  including the
  interaction associated with Zener  tuneling among the  energy levels
  pertaining  to  adjacent quantum  wells.    By  using a Monte  Carlo
  technique  we  have simulated   the  hopping motion  associated with
  absorption and emission of polar  optical phonons and determined the
  main transport parameters for the case of a GaAs/GaAlAs structure at
  room temperature.  The interaction among  the levels is found to  be
  responsible for a   systematic  increase of the  level   energy with
  respect to the  bottom of the  quantum well at electric fields above
  about 20 kV/cm.  When compared with the non-interacting case, at the
  highest fields the  average  carrier energy  evidences a  consistent
  increase  which leads to  a significant   softening of the  negative
  differential value of both the drift velocity and diffusivity.
\end{abstract}


\maketitle

\section{Introduction} 

In recent years,  superlattices  (SLs) received a relevant  scientific
and technological interest,  owing  to their nonlinear electrical  and
optical   properties.  In particular,   at increasing   electric field
strengths SLs  exhibit a  nonOhmic  behavior  with the presence   of a
negative differential conductivity (NDC) region \cite{grahn91} above a
threshold field.   In the  nonOhmic  regime, charge transport  can  be
described   in terms of   hopping  between  Wannier-Stark (WS)  states
\cite{wannier,wacker98}, thus   SLs  are good   systems  for  studying
hopping transport in semiconductor  structures.  Despite of  the above
interest, the   knowledge of   the transport  parameters,  like  drift
velocity and diffusion coefficients for a  given SL is still mostly an
open problem.
\par
From the analytical side,  the problem was investigated with different
approaches in Refs.  [\cite{rott02,bryksin03,kleinert04,  mouro,
  bonilla}].  The   coupling  between energy  levels  and the electric
field (the so called Zener  effect \cite{zener}) was found responsible
for the    onset    of   a   resonance     in  the  drift     velocity
\cite{rott02,kleinert04}      and   of    an    antiresonance  in  the
diffusion\cite{kleinert04} at fields  where both these quantities  are
decreasing functions of the applied electric field.
\par
From  the    simulative  side,  recent   Monte   Carlo  investigations
\cite{tarakonov05,rosini05}  have   analyzed     the  drift   velocity
\cite{tarakonov05}   and  both   the  drift   velocity   and diffusion
\cite{rosini05} but neglecting the coupling  between the field and the
energy levels.  Furthermore,  in Ref. [\cite{tarakonov05}]  only
general trends  are reported so   that a quantitive evaluation of  the
drift velocity is absent.
\par
Here  we  develop a microscopic  theory of  electron transport  in SLs
within the  Wannier-Stark approach, which  includes the coupling among
the energy levels due to the presence of a strong electric field so as
to cause  Zener   tunneling.  The theory    is applied to  an  ideally
infinite  GaAs/GaAlAs   superlattice at 300    K.  The electric fields
considered are in the range $5 \div 200$ kV/cm where the Wannier Stark
approach is well justified \cite{wac98, wakjah}, for which experiments
are available  \cite{madhavi02}.  A comparative  investigation of  the
effect of the interaction among energy levels with respect to the case
of independent energy levels is then carried out.
\par
The content of the paper is organized as follows.  In Sec. 2 we define
the physical system  under investigation  and present the  theoretical
approach to  be used within a Monte  Carlo simulator.  In Sec.  3 the
main results  concerning    the  drift velocity and    the   diffusion
coefficient are presented. Major conclusions are drawn in Sec. 4.
\begin{figure}[htb]
\centering
 \includegraphics*[width=.75\linewidth]{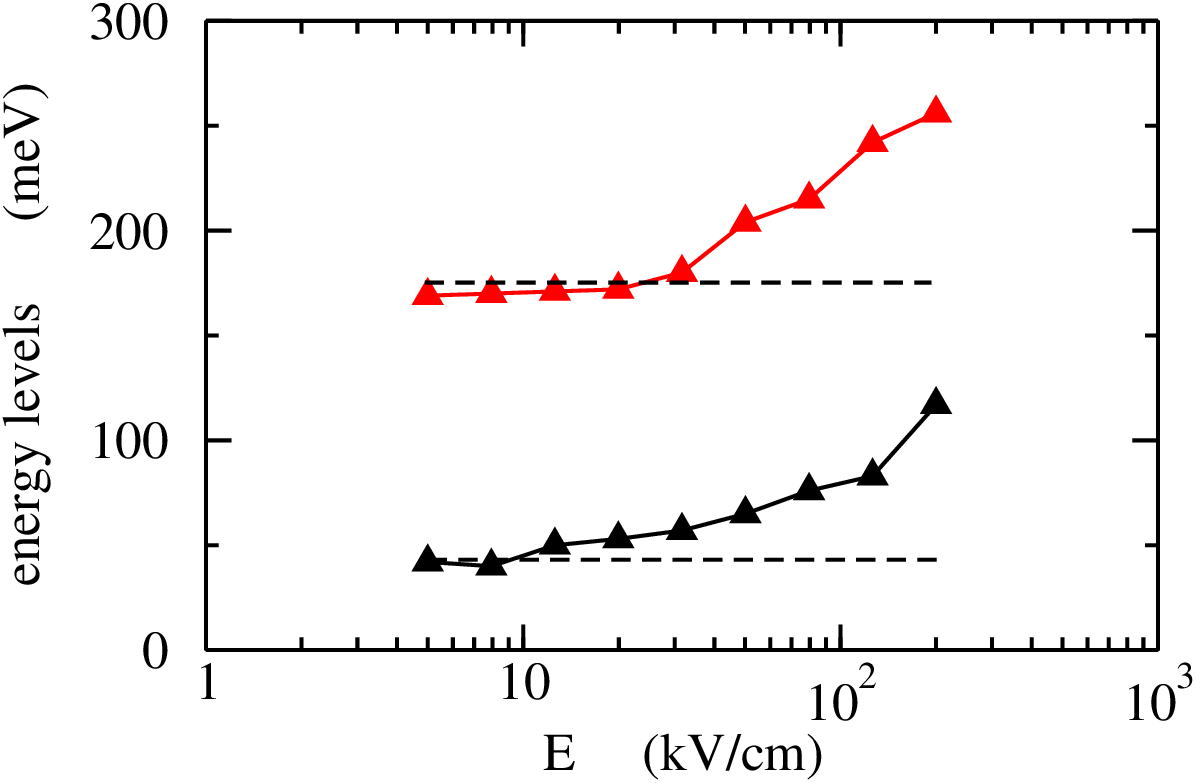}
 \caption{Eigenvalues   of  the first and  second   energy levels of a
   quantum well pertaining to an ideal infinite  SL.  Symbols refer to
   the case of interacting energy levels, dashed lines  to the case of
   independent energy levels.}
 \label{fig1}
\end{figure}
\begin{figure*}[htb]
\includegraphics*[width=.32\textwidth,height=3.5cm]{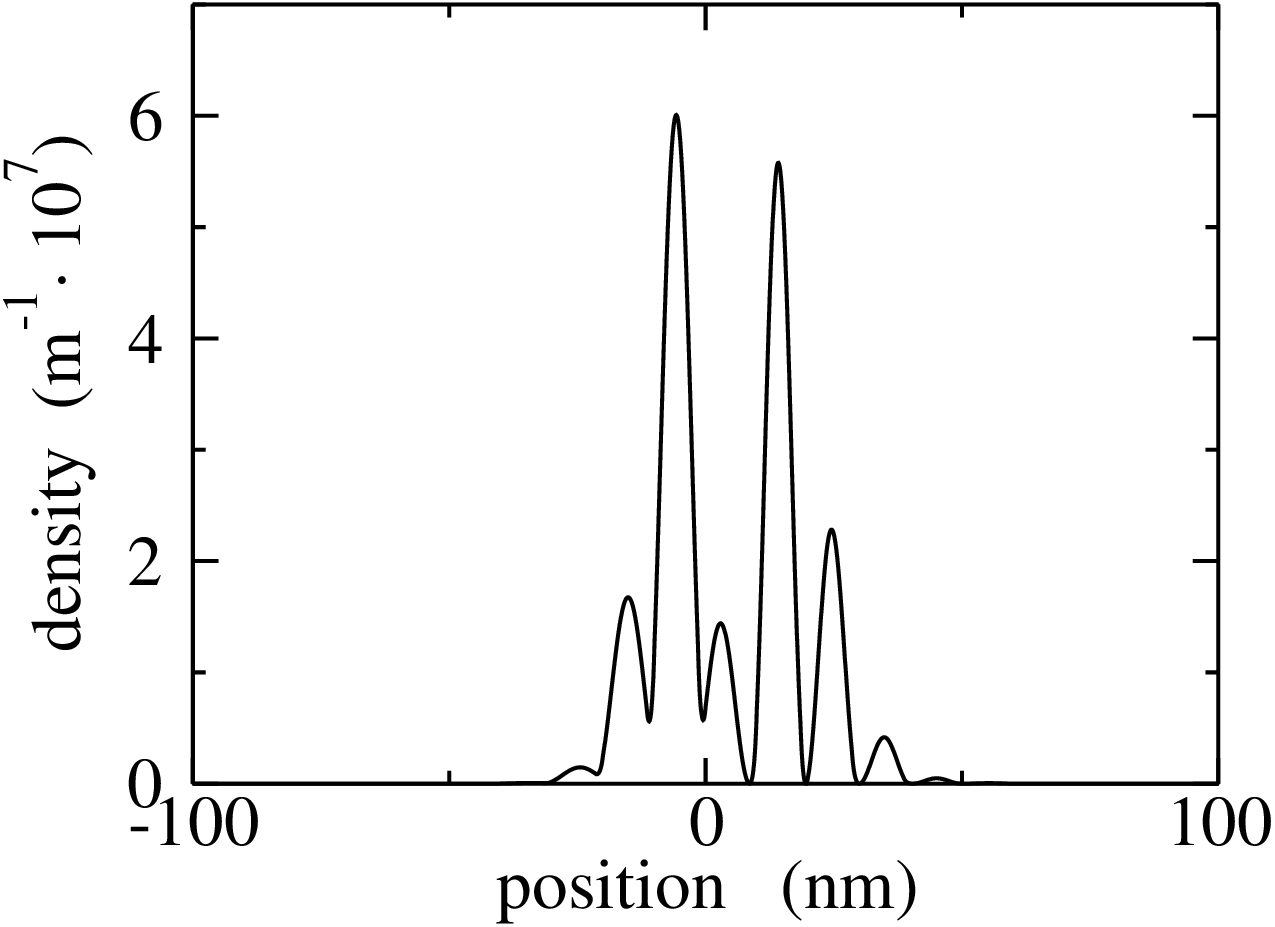}\hfill 
\includegraphics*[width=.32\textwidth,height=3.5cm]{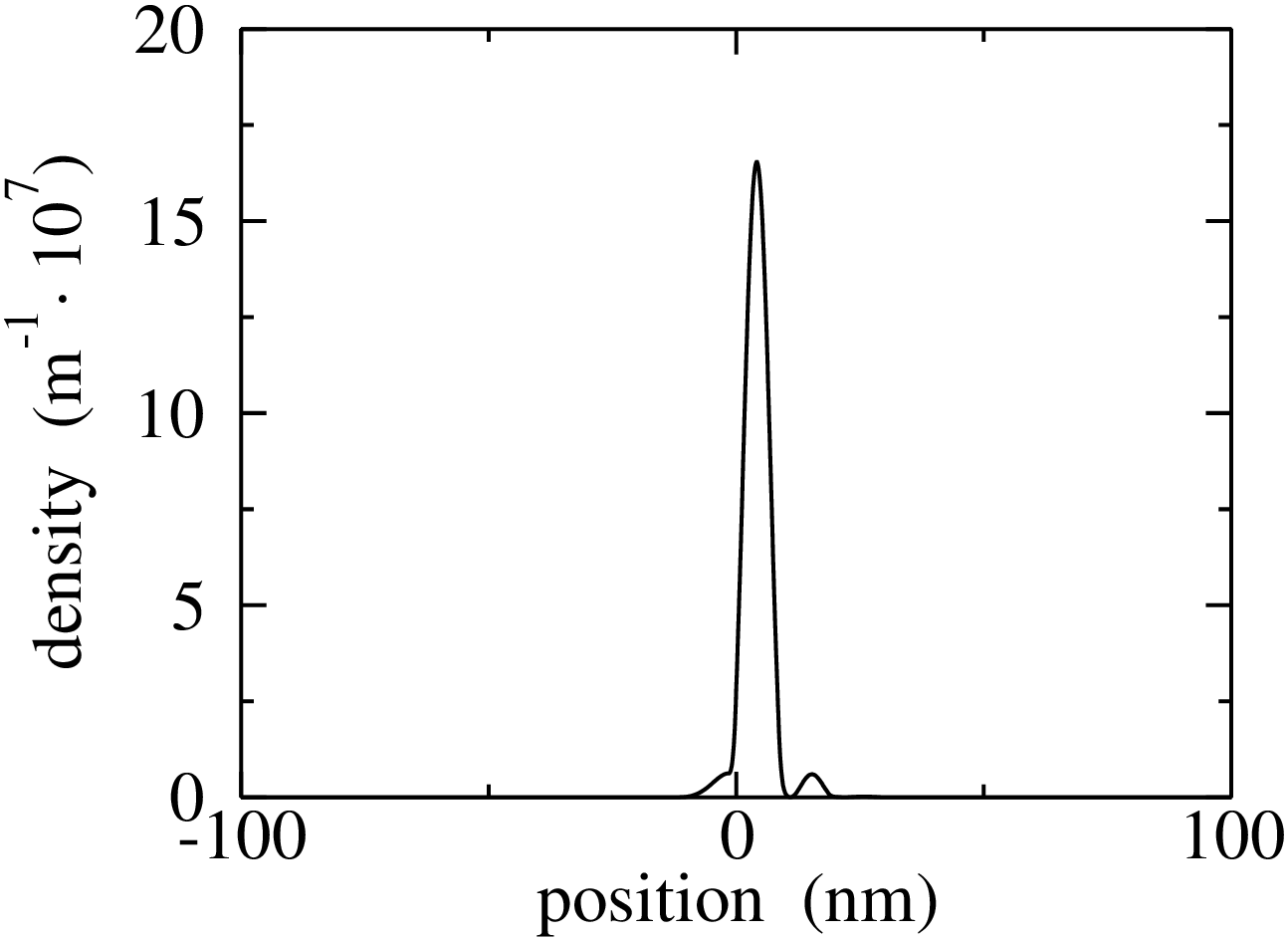}\hfill 
\includegraphics*[width=.32\textwidth,height=3.5cm]{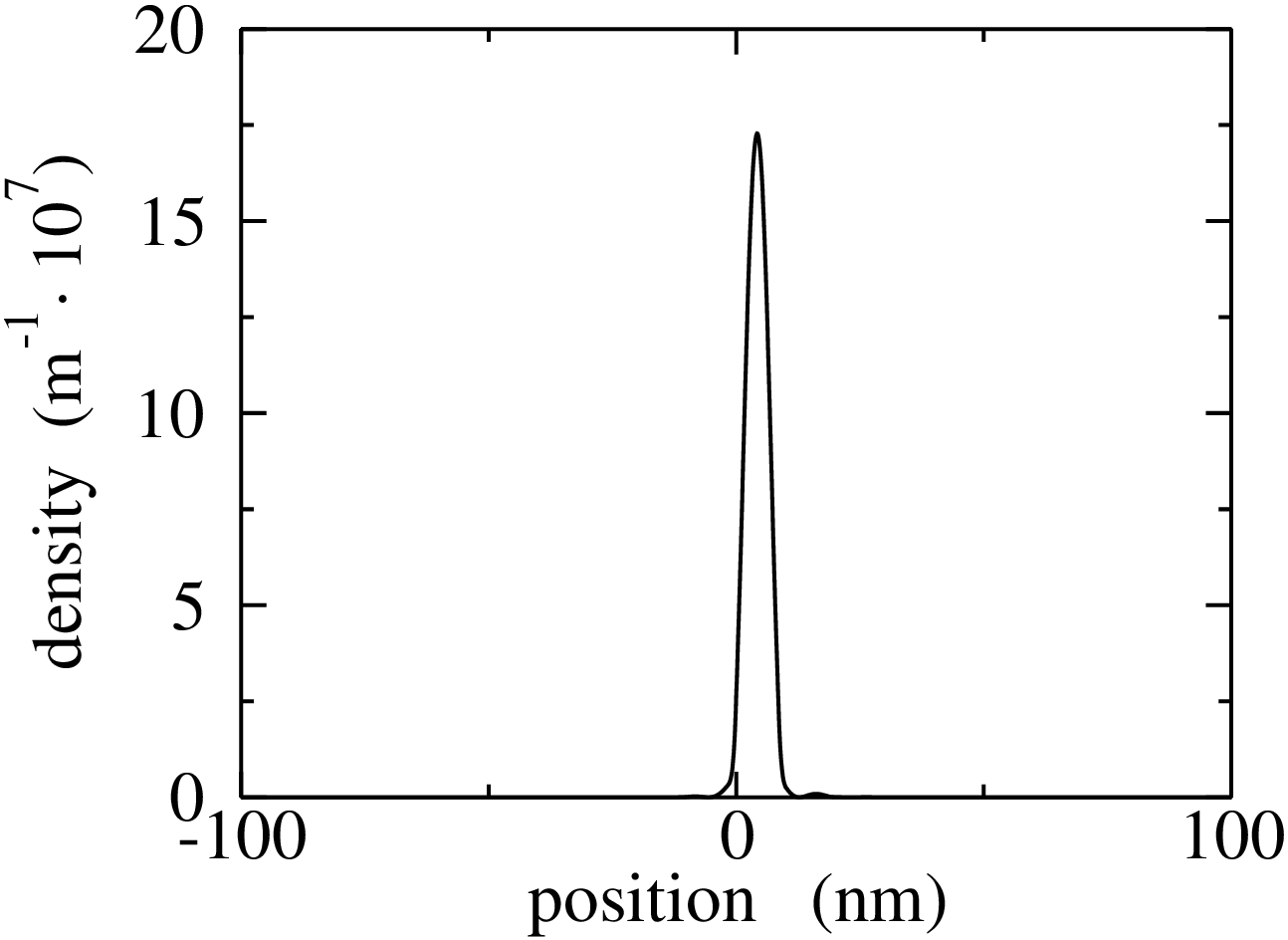}\hfill
\includegraphics*[width=.32\textwidth,height=3.5cm]{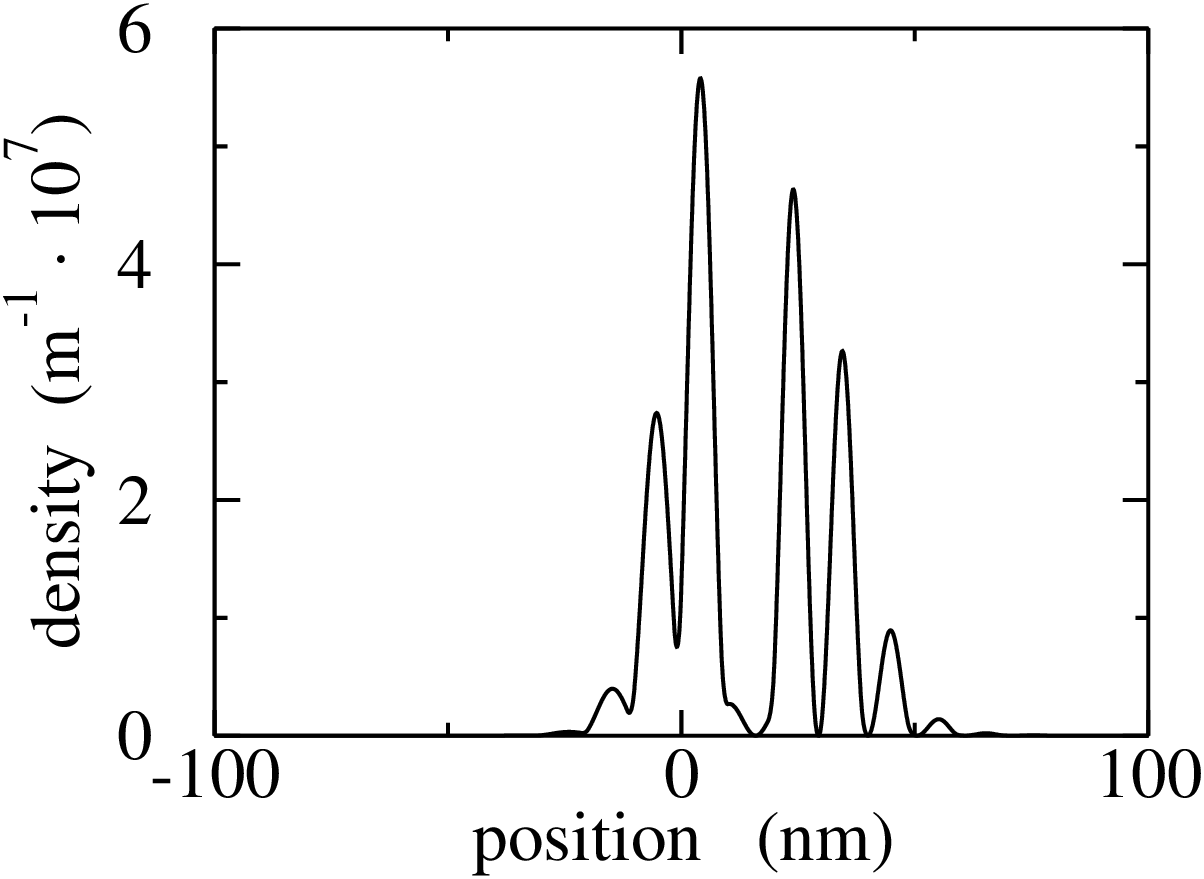}\hfill 
\includegraphics*[width=.32\textwidth,height=3.5cm]{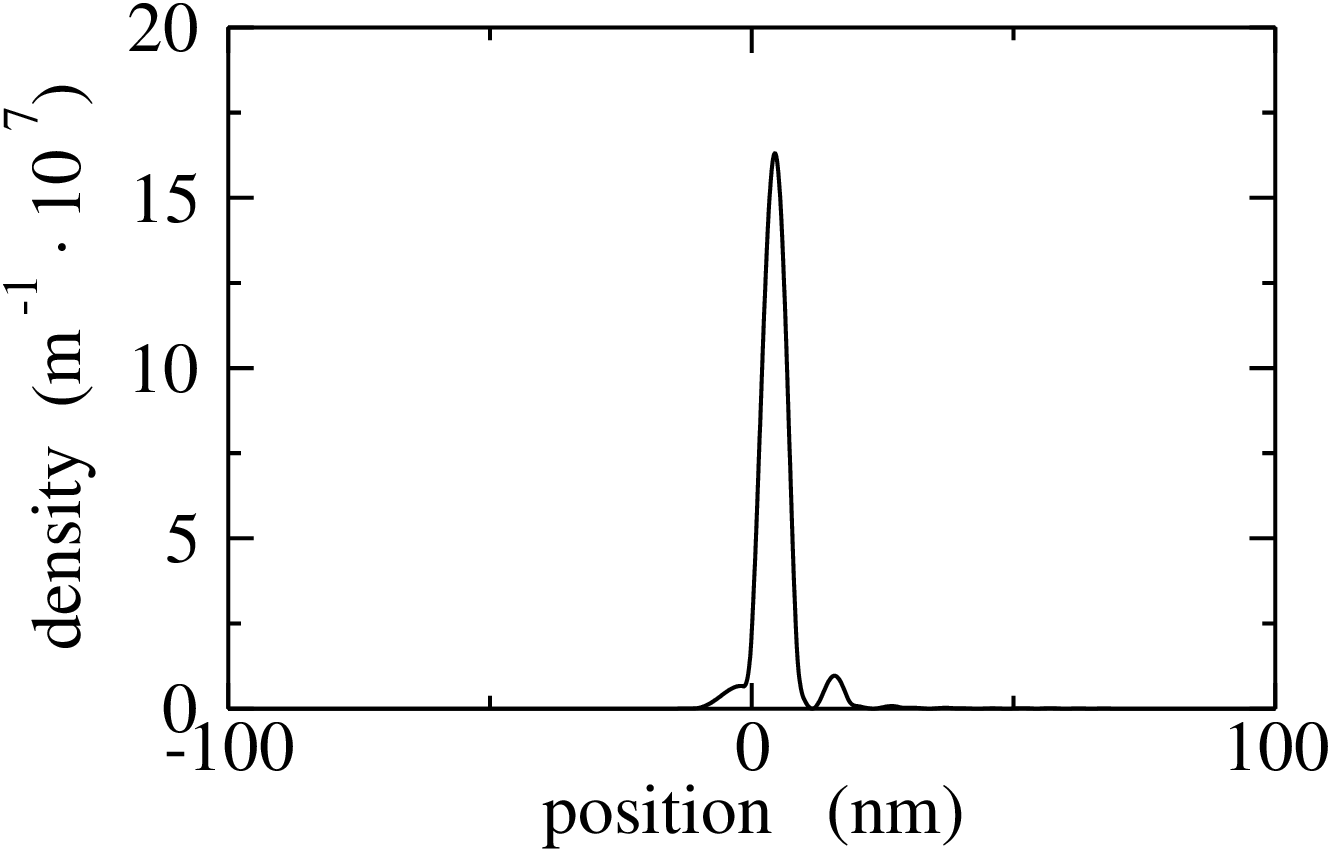}\hfill 
\includegraphics*[width=.32\textwidth,height=3.5cm]{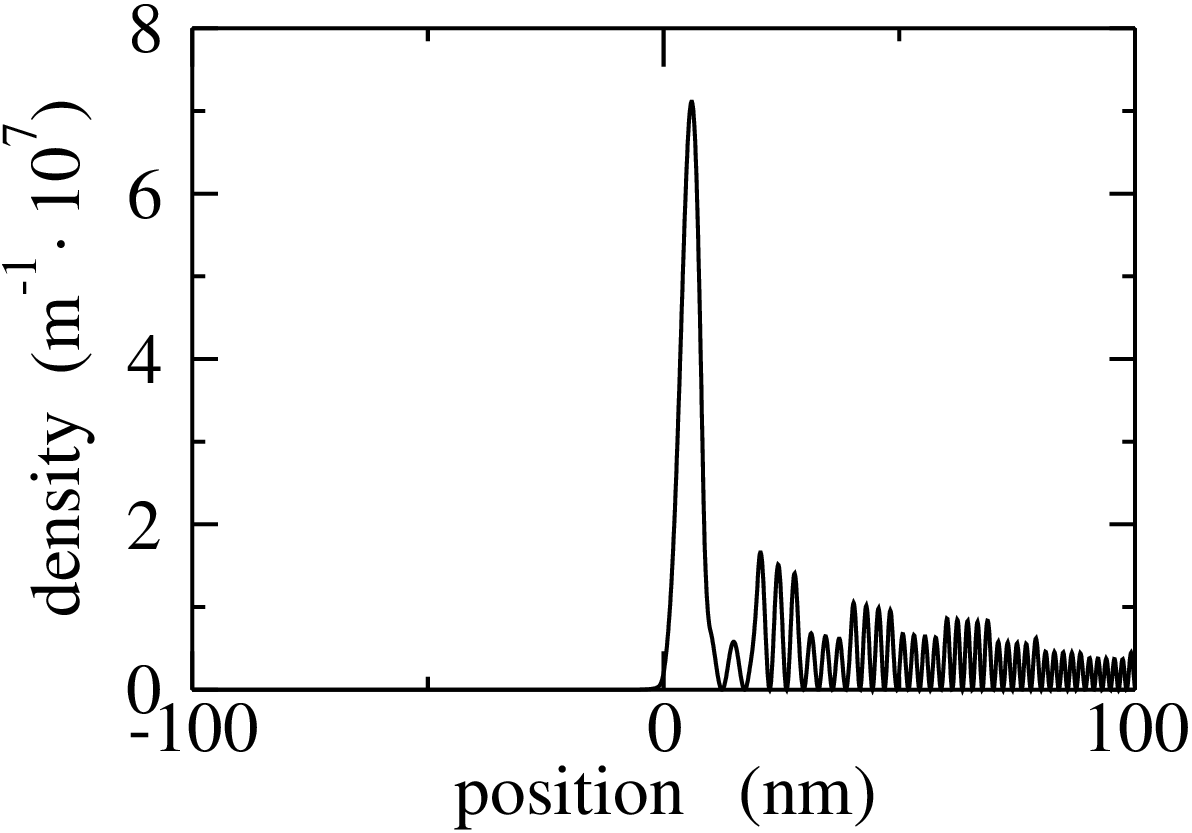}\hfill
\includegraphics*[width=.32\textwidth,height=3.5cm]{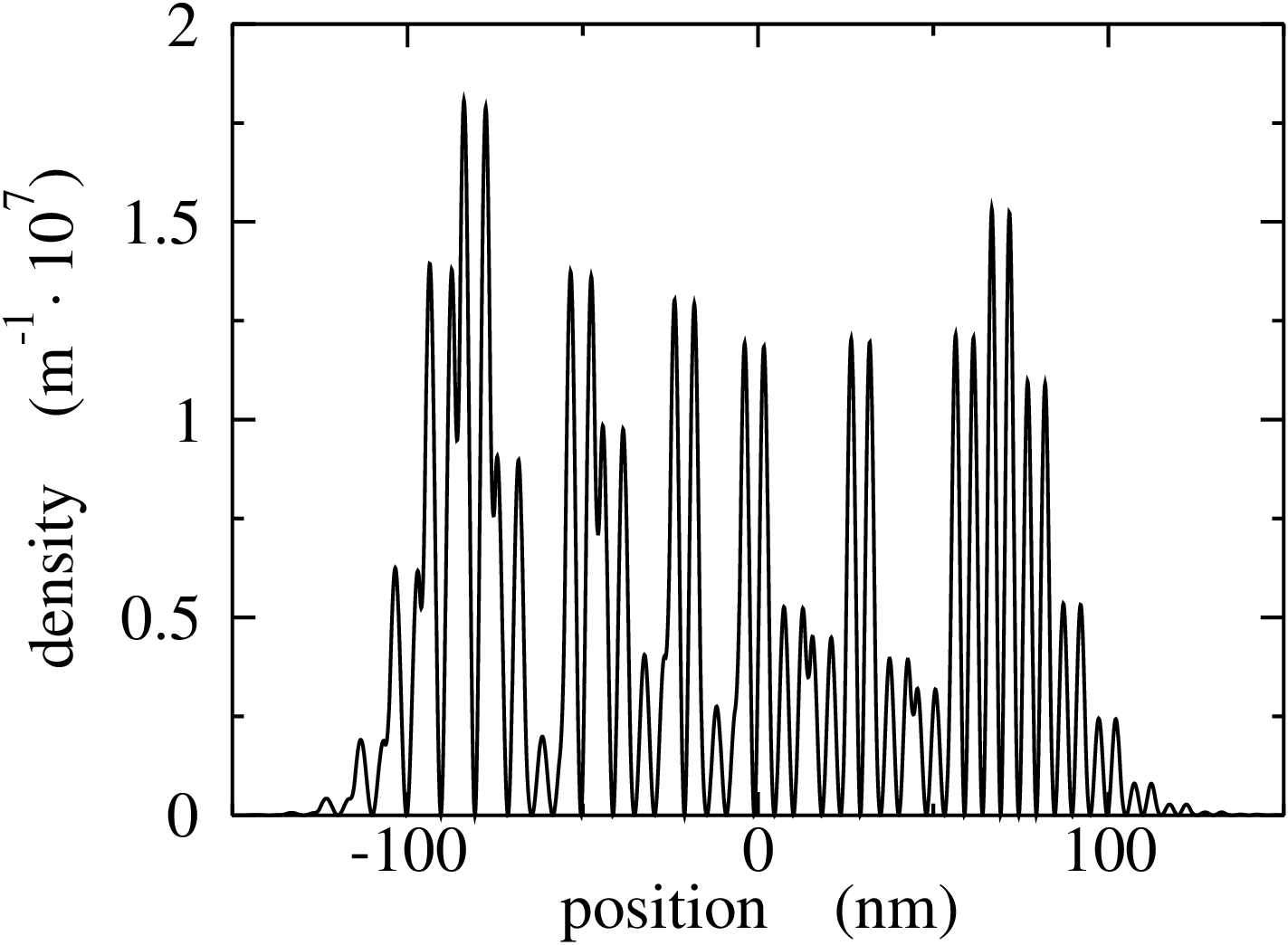}\hfill 
\includegraphics*[width=.32\textwidth,height=3.5cm]{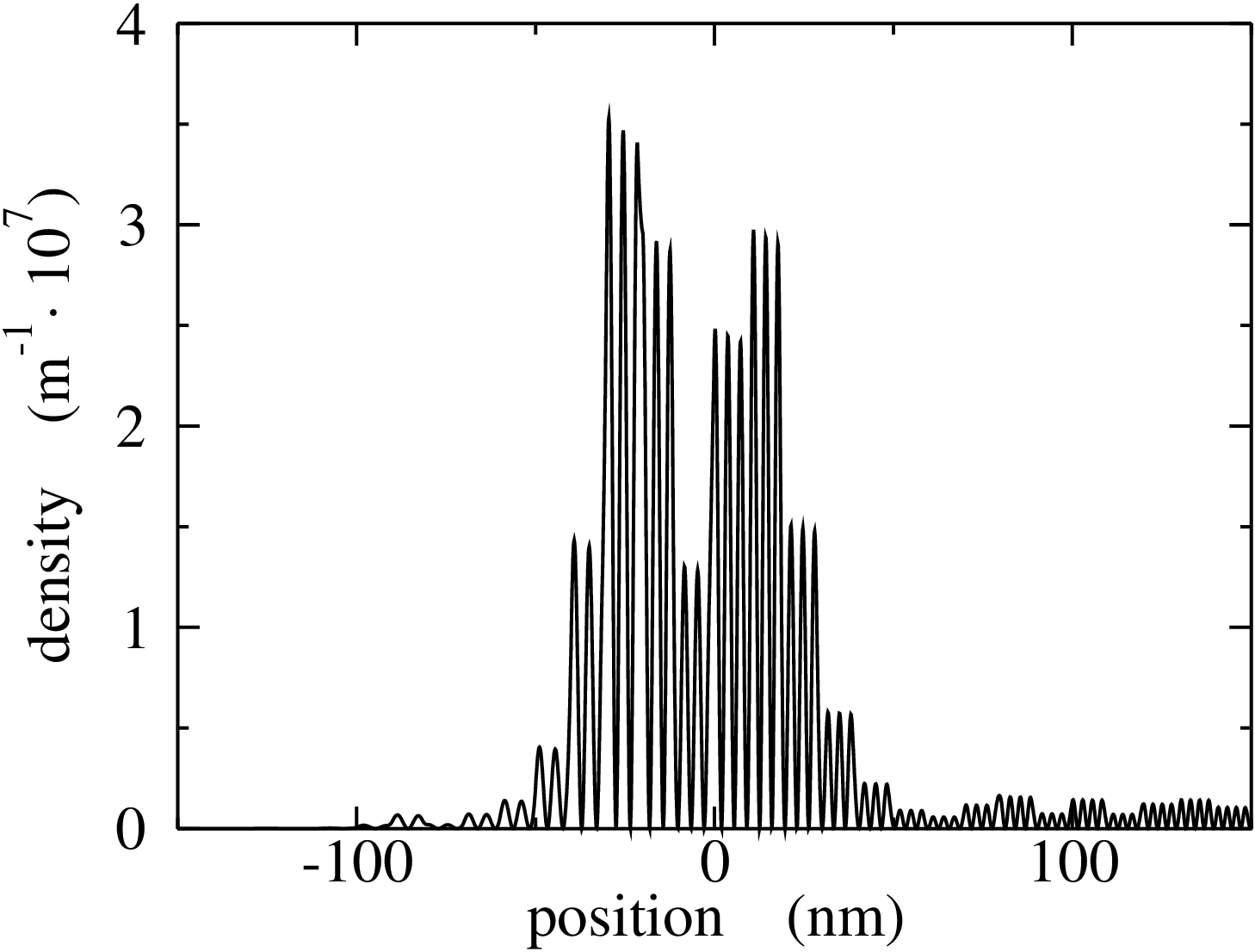}\hfill 
\includegraphics*[width=.32\textwidth,height=3.5cm]{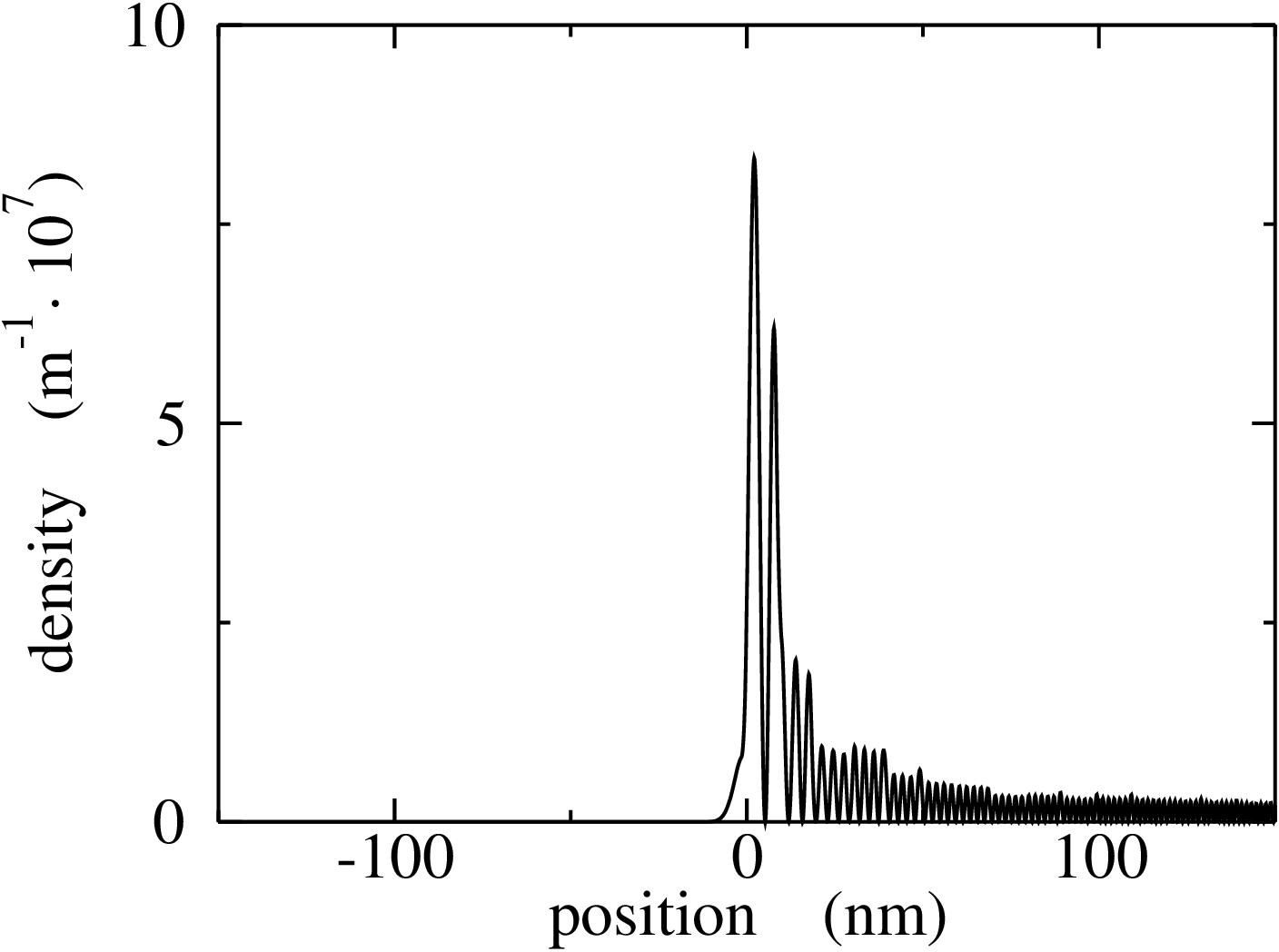}
\caption{ Normalized square value of the Wannier Stark wavefunction at
  different electric  field strengths of,  respectively  from right to
  left,  5, 30 and 200 kV/cm.  The first raw of  figures refers to the
  first level in the absence of interaction; the second raw of figures
  refers to the first level in  the presence of interaction; the third
  raw of  figures  refers  to the  second  level in  the   presence of
  interaction.  }
\label{fig2} 
\end{figure*}
\section{Theoretical approach}

\subsection{Theory of WS states}
We consider a SL made by alternated layers of different semiconductors
(A and B) along the $z$-direction.  The single electron \sch\ equation
in the presence of an external  electric field along the $z$-direction
$ \textbf{F}=(0,0,F) $ is:
\begin{equation}
\left[ \frac{\hat{\textbf{p}}^2}{2 m^*}+ V_{SL} (\textbf{r}) -
eFz\right] \psi(\textbf{r})= \varepsilon \psi(\textbf{r})
\end{equation} 
where $\hat{\textbf{p}}$   is the   momentum operator,  $m^*$   is the
effective mass assumed to be the same in  the well and in the barrier,
$V_{SL} (\textbf{r})$ is  the  potential which determines  the barrier
height of the SL, and $\psi(\textbf{r})=\varphi(z)g(x,y)$ with $g$ the
Bloch function.  The SL potential is taken of the Kronig-Penney form:
\begin{equation}
V_{SL}(z)=
   \left\{ \begin{array}{ll} 
              V_0 & \mbox{  if } 0<z<a \\
              0   & \mbox{  if } a<z<a+b
           \end{array}
   \right.
\end{equation}
with $a$  and  $b$  the  length of   the  well and  of  the   barrier,
respectively, and $d=a+b$ the SL period.  Accordingly, using the Dirac
notation the \sch\ equation in the $z$-direction is:
\begin{equation}
\label{hdirac}
[\oH_0 - eFz]\ket{\varphi}=\epsilon\ket{\varphi}\ .
\end{equation}
The solution of  Eq. (\ref{hdirac}) can  be expanded over the SL Bloch
states $\ket{n,k_z}$,  with $n$ the  band index, which are eigenstates
of the Hamiltonian $\oH_0$  of the SL in the  absence of  the electric
field
\benl
\ket{\varphi}=\sum_{n'}\int_{k_z'}\de k_z'\scal{k_z'n'}{\varphi}\ket{k_z'n'}
\ee
and  substituted  into  the    above \sch\ equation;  multiplying   by
$\bra{k_z n}$ we  develop the calculation and obtain
\bea\label{inverseh}
\left[E_n(k_z)-\epsilon\right]\scal{k_z n}{\varphi} &-& \nonumber \\
eF\sum_{n'}\int_{k_z'}\de k_z'\media{k_z n}{z}{k_z' n'}\scal{k_z'n'}{\varphi}
&=&0\ .
\eea
Here, the value  $\media{k_z n}{z}{k_z' n'}$ is calculated  explicitly
and, after some algebra, it results to be
\benl
\media{k_z n}{z}{k_z' n'}= \frac{1}{i}\deriv{}{k'_z}
\delta_{nn'}\delta{(k_z-k_z')}+X_{nn'}\delta{(k_z-k_z')}
\ee
with
\begin{equation}
\label{xnnp}
X_{nn'}=\frac{2\pi i}{L}\int\de z
u^*_{nk_z}(z)\deriv{}{k'_z}u_{n'k'_z}(z) \ .
\end{equation}
where  $u_{nk_z}$ is  the  periodic part of the  Bloch  function.  The
above matrix elements include  the interaction among levels pertaining
to different quantum  wells caused by the  electric field and they are
basically related to Zener interband tunneling.  We recall, that under
the conditions of  moderate electric fields  and weak coupling between
adjacent  wells it is justified to  consider a one-band  model for the
calculation, thus neglecting off-diagonal matrix elements described by
eq.  (\ref{xnnp}).  This approximation  leads to a simple solution  of
eq.  (\ref{inverseh})     and   was    already    applied  in     Ref.
[\cite{tarakonov05,  rosini05}]  where  the WS eigenstates  have
been obtained analytically \cite{rosini05}.

\subsection{Numerical method}
In the present   paper   we point to  fully   take  into  account  the
off-diagonal  terms (\ref{xnnp})     in the calculation    of   the WS
eigenstates and further of the   SL transport properties. However,  in
this case it  is not possible to  get a simple  analytical solution of
the system. Thus, to accomplish this purpose, we have decided to solve
numerically the \sch\ equation along the $z$ direction (\ref{hdirac}).
With  this approach, no interaction  is dropped during the calcualtion
and the interaction decribed by (\ref{xnnp}) is automatically included
in the solution.  Looking for a solution of the  \sch\ equation at the
finite differences, we  consider a number $M$  of cells (typically  30
cells in the backward direction and 30 in the  forward direction for a
total  number of $M$  =  61) representative of  an ideal  infinite SL.
(Indeed, electron wavefunctions are found to extend up to a maximum of
20 cells).  Then, the \sch\ equation is discretized in a number of $N$
points (typically  $N$= 100) equally spaced inside  a  single cell, $M
\times   N$ being  the total   number of  points   used  to obtain the
numerical solution within an estimated  convergence of a few  percent.
Accordingly,   $z \rightarrow  i  \Delta $,   $ \varphi(z) \rightarrow
\varphi_i $, with $ \Delta=\frac{d}{N} $, and defining
$$
d_i= \frac{\hbar^2}{m^*} \frac{1}{\Delta^2}+V_i-eFi\Delta 
$$
$$
e=-\frac{\hbar^2}{2m^*} \frac{1}{\Delta^2}
$$
$$
\textbf{A}=\left( 
\begin{array}{cccccc}
d_1 & e & 0 & \ldots & \dots & 0  \\
e & d_2 & e & 0 & \ldots  & 0\\ 
0 & e & d_3 & e & \ldots & 0 \\
\vdots & {} & \ddots & \ddots & \ddots & \vdots \\
0 &\ldots & 0 & e & d_{N M-1} & e \\
 0 & \ldots & \ldots & \ldots & e & d_{N M}
\end{array}
\right)
$$
the discretized \sch\ equation of the system takes the compact form:
\begin{equation} 
\textbf{A} \varphi = \varepsilon \varphi 
\end{equation}
with $\varphi = (\varphi_1,\varphi_2,\ldots,\varphi_{NM})$.
\par
Since the matrix $\bf  A$  is real and  symmetric the  eigenvalues are
real, distinguishable, and in number equal to  the order of the matrix
$ M \times N$.    The eigenvectors $\varphi^i$ are  the  Wannier-Stark
wavefunctions.   In  the present   case,  the  $  \varphi_j^{i}  $ are
two-dimensional  vectors  with    the index    $i$   representing  the
corresponding   energy level  and  the index  $j$  the pertaining grid
point.  The  eigenvalue   $ \varepsilon^i  $  and  the eigenvector   $
\varphi^i $,   being labelled with the   same index $i$,  are  in full
correspondence.
\par
The   eigenvalues are then  classified  according   to the  pertaining
quantum   well and   energy level   by  using the  properties  of  the
Wannier-Stark eigenvectors (for  a    review  of the  WS     functions
properties see for example \cite{wakrep}).  In particular we know that
WS  functions should be     localized and satisfy  the   translational
symmetry $\varphi_{n+m}(z)=\varphi_n(z-md)$  for any pairs  of quantum
wells labelled by  $(n,m)$.  Then, as validity  test of the  numerical
algorithm, the energy  difference between adjacent wells is calculated
and their ladder distribution verified. We  have also verified that in
the central  region    of the representative   device,  where boundary
effects are negligible, the translational property of the eigenvectors
belonging to the same energy level  is accomplished.  In this way, for
each   value   of  the   electric  field we    select the wavefunction
representative of the infinite SL.
\subsection{Electron-phonon interaction}
The calculation of the polar-optical phonon scattering probability per
unit time in the  WS representation follows  the standard Fermi Golden
Rule.  In brief, after decoupling the  phonon bath, the point-to-point
transition probability per unit time is
\bea\label{finalprob}
&&P(\nu n \kv, \nu' n' \kv') = \frac{2\pi}{\hbar}\sum_{\qv}c^2(\qv)
\left[\ba{cc} n_{\qv} \\ n_{\qv}+1 \ea\right]\times
\nonumber \\
&&\left|\bra{\nu' n'}e^{\mp iq_z z}\ket{\nu n}\right|^2 \times
\nonumber \\
&& \delta_{\kv-\kv'\mp\qv_{\parallel}}
\delta\left[\epsilon(n\kv)-\epsilon(n'\kv')
  \mp\hbar\omega_{op}+\nu eFd\right]
\eea
where  $n_{\qv}$ is  the mean value   of the phonon occupation number,
$\qv=(\qv_\parallel,q_z)$     is   the phonon      wave  vector    and
$\boldsymbol{\mathrm  k}=(\kv,k_z)$.   For   $c^2(\qv)$ we   take  the
standard Fr\"olich squared matrix element \cite{frol}:
\be{hamf1} c^2(\qv) = \frac{\hbar e^2
  \omega_{op}}{2V\epsilon_0}
\left(\frac{1}{\epsilon_+}-\frac{1}{\epsilon_-}\right)
\frac{\qv^2}{(\qv^2-\qv_D^2)^2}
\ee
where $\omega_{op}$   is the  optical phonon frequency,  $\epsilon_-,\
\epsilon_+$ are  the static  and   high frequency relative  dielectric
constant, $\qv_D$ is the screening wave vector and  $V$ is the crystal
volume.
\par
Being  interested  in the total  scattering  probability per unit time
from  an initial  state  $(\nu  n \kv)$ to  any   possible final state
$\kv'$, we  integrate over the  final momentum variable and obtain the
integrated transition probability per unit time:
\bea
\label{scatfi}
&&P_{\nu'n'}(\nu n\kv)=\frac{m^*}{(2\pi)^5\hbar^3}
\left[\ba{cc} n \\ n+1 \ea\right] \times
\nonumber \\
&&\int{\de \theta}{\int\de q_z}
c^2(\qv)\left|\bra{\nu' n'}e^{\mp iq_z z}\ket{\nu n}\right|^2\ .
\eea
Having  integrated  the   $\delta$ function  in  Eq. (\ref{finalprob})
implies  having accounted for momentum conservation  that leads to the
constraint   $\qv=(k'_x-k_x,\   k'_y-k_y,\  q_z)$.    This   last  Eq.
(\ref{scatfi}) is not further  integrable with standard method because
of the  not simple dependence  upon  $\qv$.  Within  the  Monte Carlo
technique  this step  is accomplished  using  the rejection  technique
\cite{jacoboni83}.
\par
Equation  (\ref{scatfi}) describes  the  probability that  an electron
with initial state $(\nu n\kv)$ is scattered  to a final state in well
$\nu'$, in band  $n'$, with   any  $\kv'$.  However, in our   approach
electrons  are allowed to  jump a maximum  distance of 6 wells forward
and  6  wells backward,  since   the  hopping  transition  element  is
negligible for longer distances.  Indeed,  we have verified that,  for
the  transport parameters   investigated  here,  the results are   not
affected by  a   further increase  of the   maximum  hopping distance.
Accordingly, the associated transition probabilities per unit time for
a carrier jump from an  initial well to any of  the 12 final wells due
to absorption and emission  of polar optical  phonons is evaluated and
used   inside  a  Monte   Carlo   procedure,    as detailed  in   Ref.
[\cite{rosini05}].  For sake of simplicity, other scattering
mechanisms are neglected and we have considered at most only the first
two  energy levels  in each  well.  The time evolution  of the carrier
ensemble  is then evaluated for a  period sufficiently long to extract
the drift velocity and the longitudinal diffusion coefficient from the
corresponding slopes of the average  value of the  position and of its
variance along the field direction \cite{rosini05}.
\section{Results and discussion}
The theory is applied to the case of a  GaAs/GaAlAs SL at 300 K, being
that the most  studied SL.  As typical  parameters we take $a=8.5$ nm,
$b=1.5$ nm and a barrier height of 0.4 eV.  For the electron effective
mass  we take everywhere the  value  of GaAs  $m^*=0.067 \ m_0$,  with
$m_0$ the free  electron mass, for the polar  optical phonon energy we
take $\varepsilon_{op}=35$ meV, and for the  polar optical coupling we
take $\frac{1}{\epsilon_+} -\frac{1}{\epsilon_-}=0.013$.  The electric
fields considered are  in  the range   $5 \div  200$ kV/cm  where  the
Wannier Stark approach is well justified \cite{wakjah}.

An ensemble   Monte  Carlo simulation has  been  set-up   in  order to
simulate  electron  transport   transport in   this  superlattice,  as
detailed   in  [\cite{rosini05}]  and   the   estimated physical
quantities are calculated as follows
\bea
\langle v_z \rangle & = & \frac{1}{N\delta t}\sum_{i=1}^{N}\delta z_i \\
D_z & = & \frac{1}{2}\frac{\delta\langle(\Delta z)^2\rangle}{\delta t} \\
\langle  \epsilon \rangle  & = &  \frac{1}{N}\sum_{i=1}^{N} \epsilon_i
\eea
for   drift   velocity,   diffusion   coefficient   end    mean energy
respectively,  where $z_i$ and  $\epsilon_i$ are the  position and the
energy of the $i$-th particle.
\begin{figure}[t]
\includegraphics*[width=\linewidth,height=.6\linewidth]{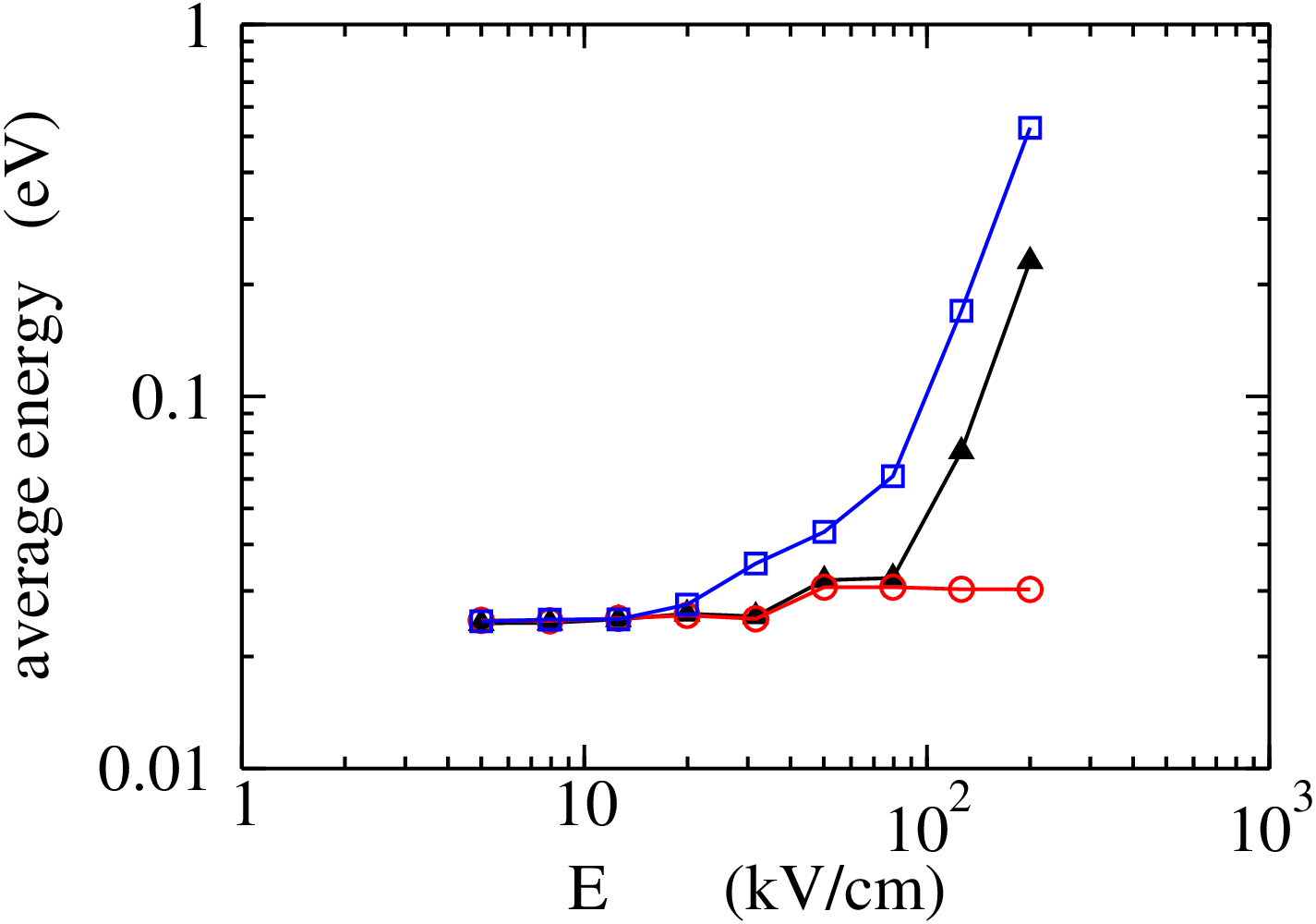}
\caption{ Average carrier energy  as function of the applied  electric
  field.  Open circles, full triangles,  and open squares refer to the
  cases when only the first level in the  absence of interaction, only
  the  first level in the presence  of interaction, both the first and
  second  level in the presence of  interaction are considered. Curves
  are guides to eyes.  }
  \label{fig3}
\end{figure}
\begin{figure}[t]
\includegraphics*[width=\linewidth,height=.6\linewidth]{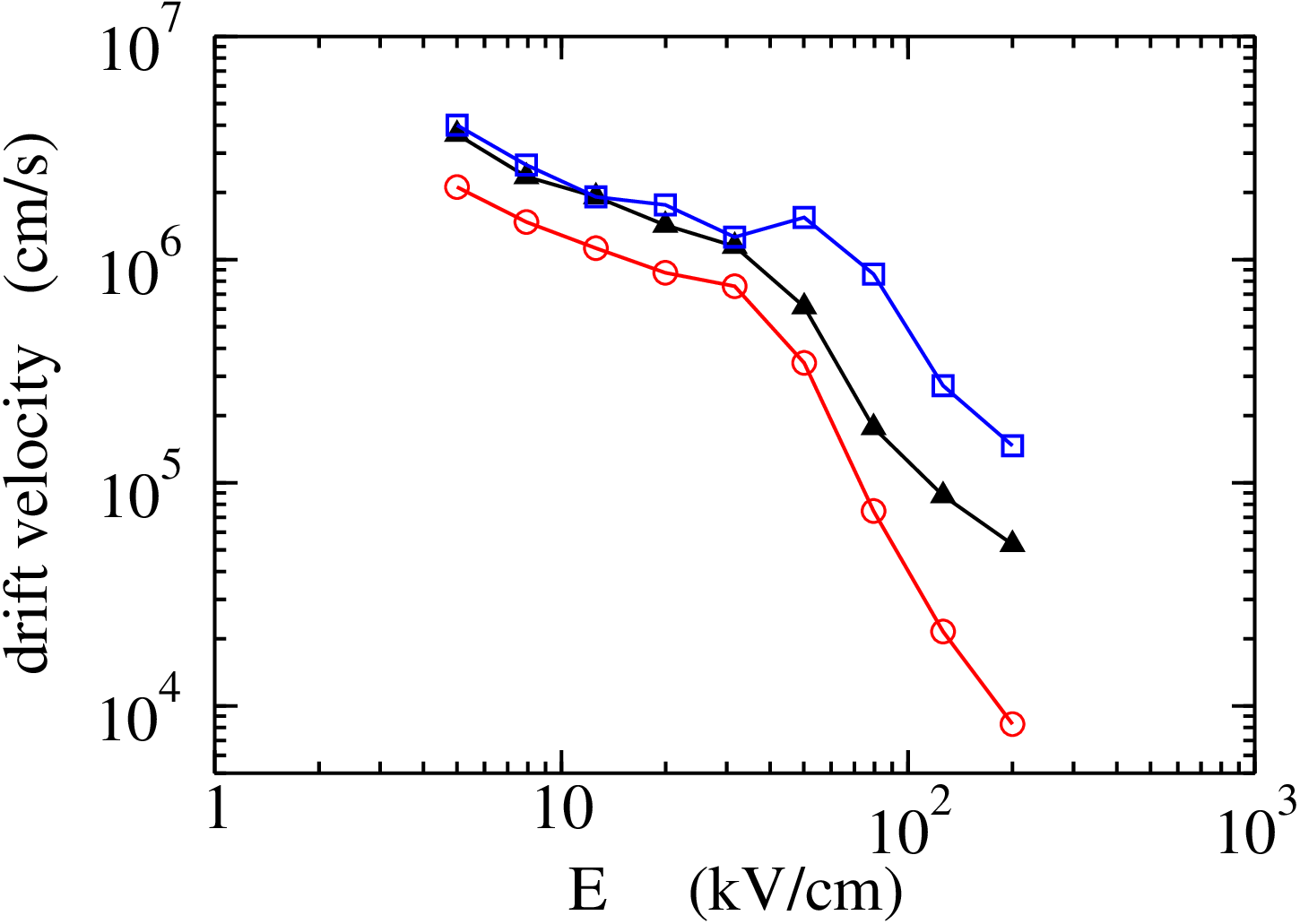}
\caption{  Average carrier velocity as   function of applied  electric
  field. Full triangles,  open circles and  open squares refer  to the
  cases when only the first level in  the absence of interaction, only
  the first level in  the presence of  interaction, both the first and
  second level in the presence  of interaction are considered.  Curves
  are guides to eyes.  }
\label{fig4}
\end{figure}
\begin{figure}[t]
\includegraphics*[width=\linewidth,height=.6\linewidth]{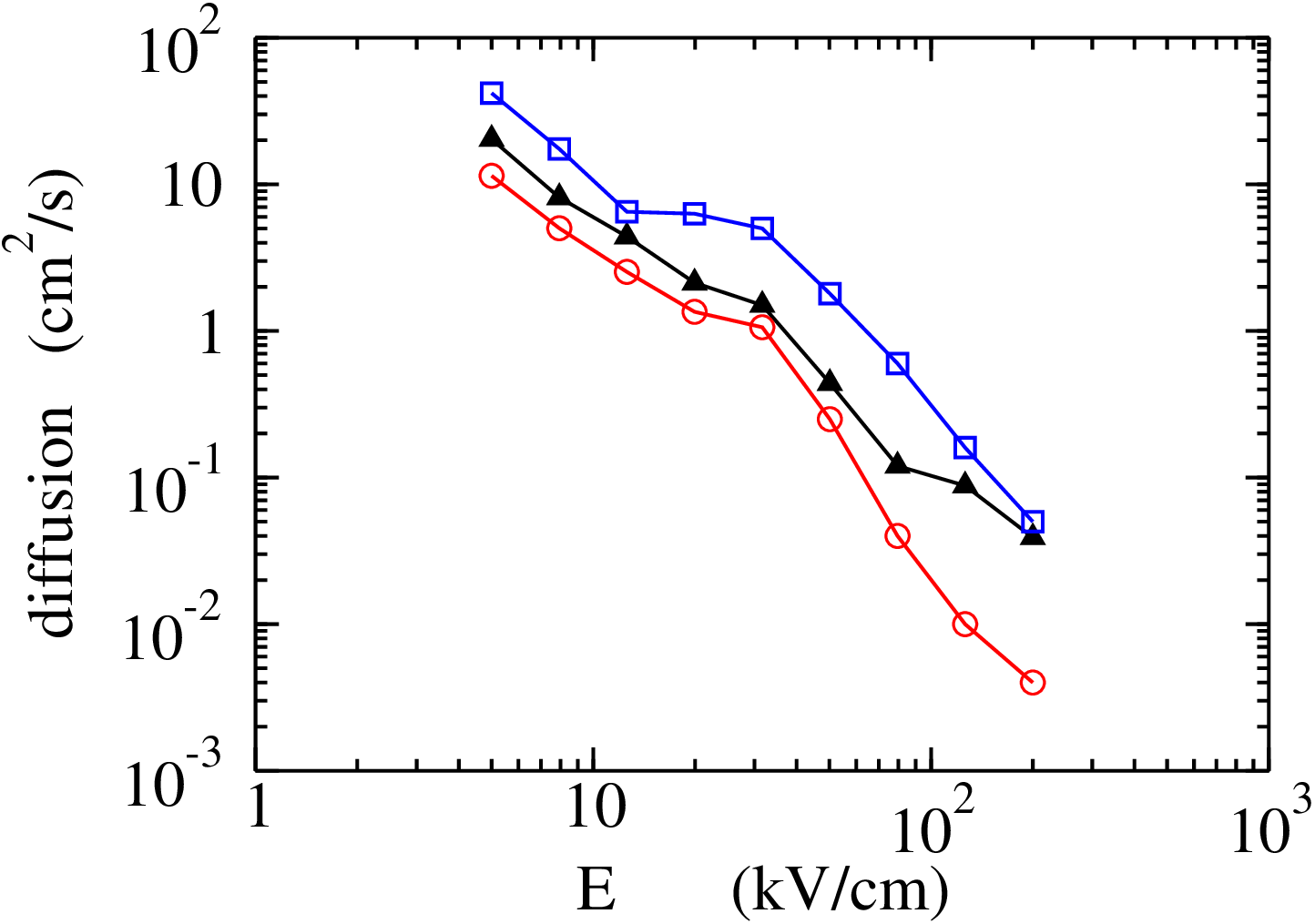}
\caption{ Carrier longitudinal  diffusion  coefficient as function  of
  the applied  electric field.  Full triangles,  open circles and open
  squares refer to the cases when only the first  level in the absence
  of interaction, only the first level in the presence of interaction,
  both the first and second  level in the  presence of interaction are
  considered. Curves are guides to eyes.  }
  \label{fig5}
\end{figure}
\par
The results of the investigation are reported in Figs.  1 to 5.  Here,
the case of interacting  levels (through Zener tunneling)  is compared
to the case of independent levels at increasing values of the electric
field.
\par
Figure 1 reports the  energy of the first and  second level inside the
quantum well (measured from the bottom of the well) as function of the
applied field.  In the absence of interaction the  energy of the first
and second levels are  found to be  of 43  and 175 meV,  respectively.
The presence  of  the interaction   is responsible for  a  significant
increase,  up to about  a factor two, of  the  energy levels at fields
starting from about 10 kV/cm.  Such an increase is associated with the
broadening of energy levels \cite{nakayama91,glutsch04}.
\par
Figure 2 reports the square modulus of the Wannier Stark wavefunctions
associated with    the first and  second  energy   level at increasing
electric fields (from  left to right).  The set  in the first (second)
raw  of  the figure refers  to independent  (interacting) first energy
levels.    The tendency to   localization  exhibited by  both cases at
moderately   increasing fields is  found to  be  broken at the highest
fields  by the presence  of the  coupling  between level and field, as
evidenced by the long tail in the direction  of the field exhibited by
the wavefunction in  the second raw.   This tail is originated by  the
coupling  of   different bands of  the   superlattice induced   by the
electric field, as  described in Eq.  (\ref{xnnp}), and enables  us to
evaluate the effect  of carrier tunneling from a  well to those in the
forward direction (Zener effect  \cite{zener}).  The agreement between
analytical and numerical results at  moderate electric fields is taken
as a validation test of the numerical algorithm used here.  The set in
the third raw   of Fig. 2 refers  to  the case  of interacting  second
energy  levels.  Here, at  low  fields the  delocalization is stronger
than in the case of  the first level,  as expected for a higher energy
level. At increasing fields we observe an increase of localization and
the appearance of the long tail in the direction of the field.
\par
Figure 3 reports the carrier  average energy obtained from Monte Carlo
simulations at  increasing    electric fields.  Above   50  kV/cm, the
presence of interacting energy levels is found to be responsible for a
substantial increase  of the average energy  (up to about a  factor of
10) associated with the increased spatial mobility of the carriers due
to  the Zener effect.  The  inclusion  of the  second  energy level is
found to emphasize the trend of energy increase..
\begin{figure}[t]
\includegraphics*[width=\linewidth,height=.7\linewidth]{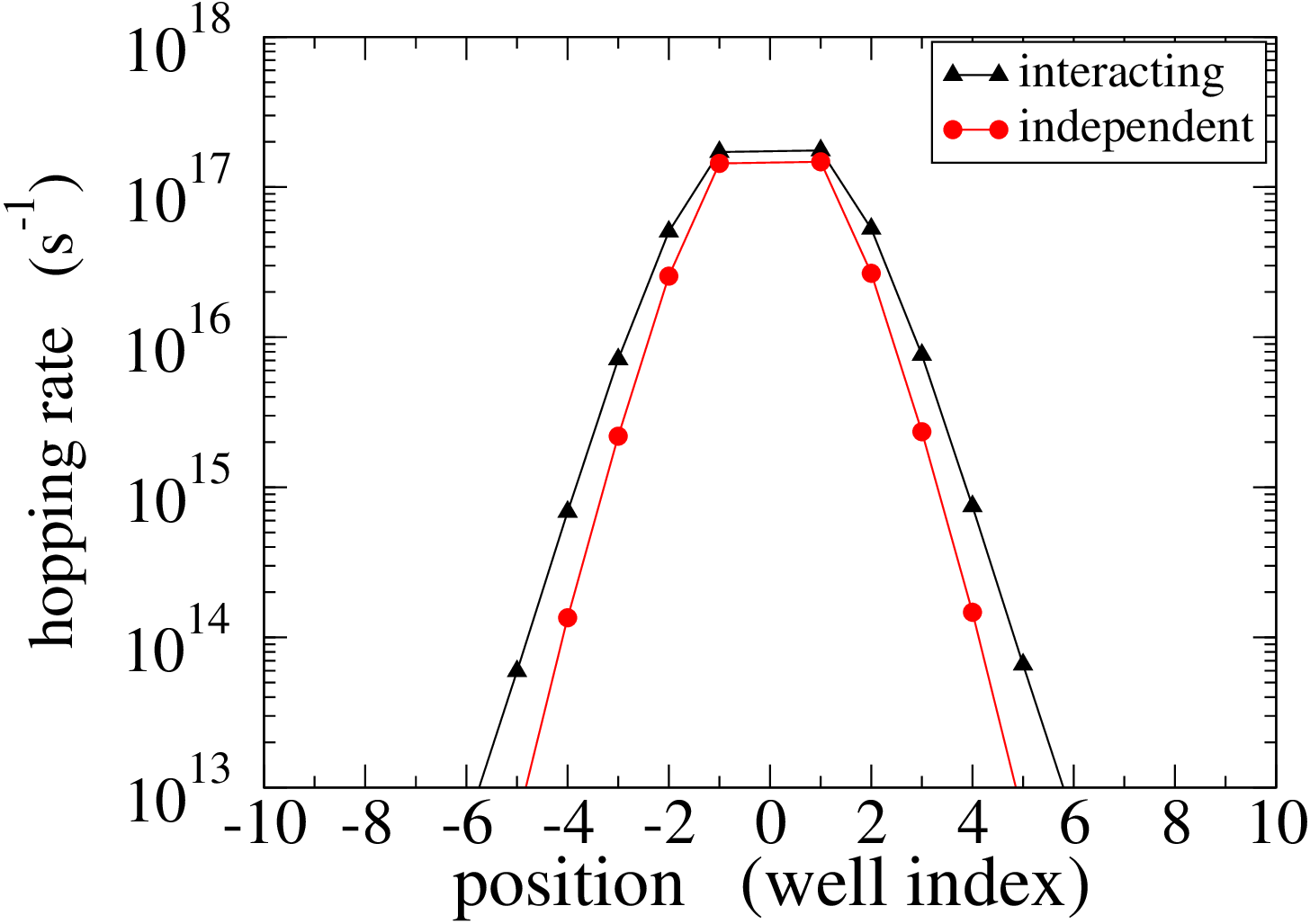}
\caption{ Maximum hopping rate  from a given inital  well $\nu=0$ to a
  final well   in   the  forward (positive)   and   bacward (negative)
  direction.  Calculations are performed    for the case of  a  single
  level   and $E=  13 \ kV/cm$.     Full circles refer to  independent
  levels, full triangles to interacting  levels.  Curves are guides to
  eyes.  }
  \label{fig6}
\end{figure}

Figures   4 and 5 report   the  average velocity  and the longitudinal
diffusion coefficient  at  increasing fields.  Apart from  a numerical
off-set within  a factor of two, in  the region up  to about 50 kV/cm,
both quantities evidence the same  systematic decrease with increasing
fields.  Such a decrease is  associated with the localization trend of
the wavefunction \cite{esa70} and it  is called negative  differential
mobility  (NDM).  Above 50 kV/cm,  the  simulations show a substantial
increase (up to about a factor of 10) of both drift and diffusion with
respect to the  independent  case, in concomitance   with that of  the
average energy.  This softening  of the negative diferential behaviour
of both drift and diffusion is associated with  the onset of the Zener
effect.    The inclusion   of the  second   energy level  is  found to
emphasize the trend to increase of  both the kinetic coefficients.  We
remark that simulations evidence an appreciable kink of both drift and
diffusion at a field around $40 \ kV/cm$, which is associated with the
resonance occurring when the potential  energy due to the voltage drop
on a period equals the optical phonon energy\cite{rott02,rosini05}.
\par
For a microscopic interpretation of the above results, we refer to the
shape of  the total hopping  rate from an  initial well $\nu$ to other
final wells $\nu'$ for a forward or backward hop in the simple case of
a  single level, $P_{\nu \nu'}$.   The  maximum value of these  rates,
which  are used in  Monte Carlo simulations to  generate a hop between
states, are reported in Figs. 6 to  8 at different field strengths for
the case of independent and interacting first energy levels.

\begin{figure}[t]
\includegraphics*[width=\linewidth,height=.7\linewidth]{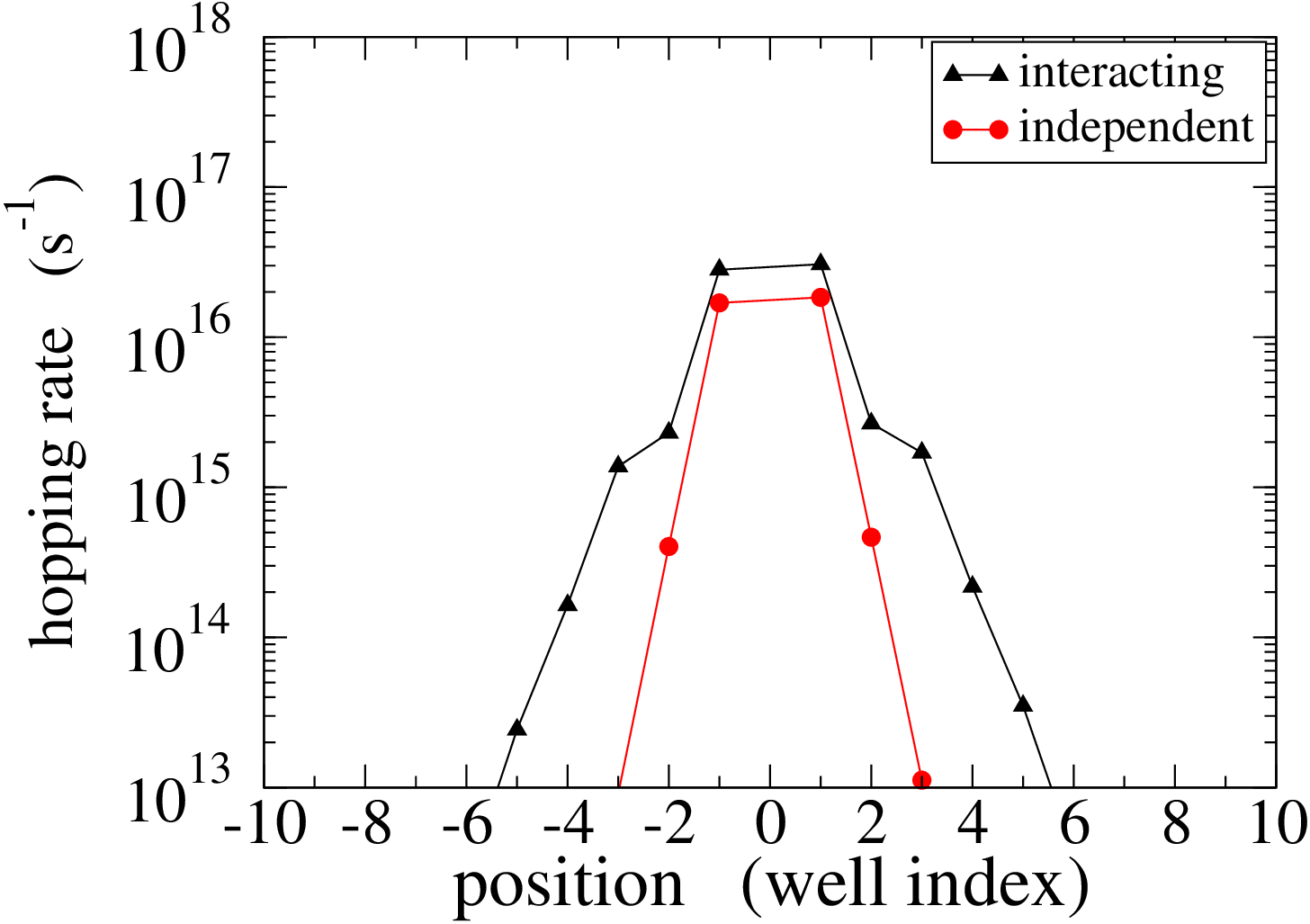}
\caption{ Maximum hopping  rate from a  given inital well $\nu=0$ to a
  final  well  in  the  forward   (positive) and   bacward  (negative)
  direction.   Calculations  are performed for the    case of a single
  level and  $E=  50  \  kV/cm$.  Full   circles  refer to independent
  levels, full triangles to  interacting levels.  Curves are guides to
  eyes.  }
  \label{fig7}
\end{figure}
Figure 6   shows that at low fields   the independent  and interacting
models  give practically the same results  (we notice that the forward
value  is always  greater than the   corresponding bacward value, thus
ensuring  a net   motion of  the   carrier  ensemble  along the  field
direction).   Hopping    rates  at increasing    distances   decreases
exponentially  as expected by the   localized feature of the  electron
wavefunction.
Figures 7 and 8 show  that the presence  of the interaction leads to a
systematic increase of the rates expecially of  those to far neighbour
wells.  Accordingly, the  drift   and  diffusion in the  presence   of
interacting levels increases with  respect to the case of  independent
levels.

\begin{figure}[t]
\includegraphics*[width=\linewidth,height=.7\linewidth]{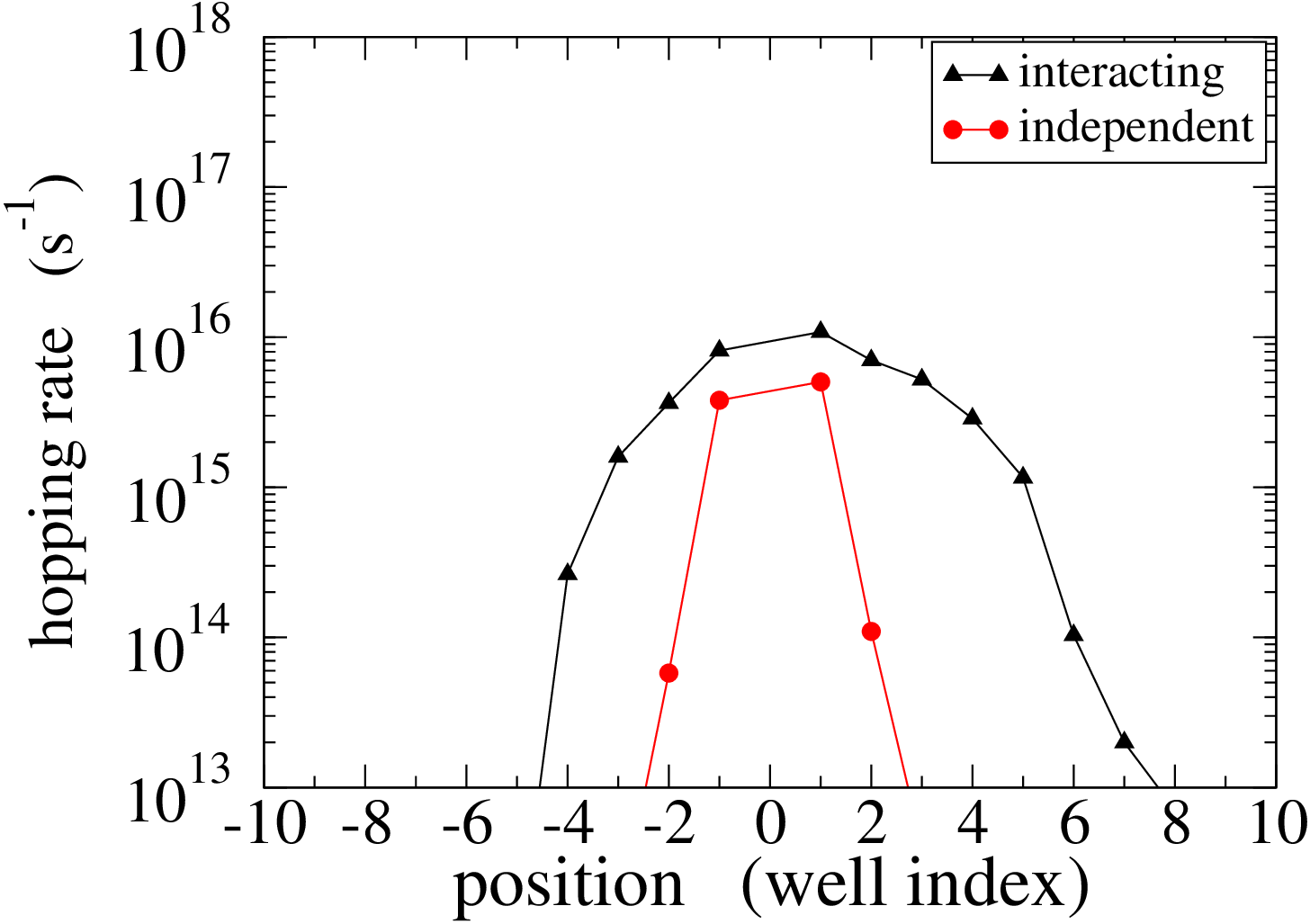}
\caption{ Maximum hopping rate  from a given inital  well $\nu=0$ to a
  final  well  in    the forward  (positive)  and   bacward (negative)
  direction.  Calculations  are performed  for the  case  of a  single
  level  and  $E= 200 \ kV/cm$.     Full circles refer  to independent
  levels, full triangles to  interacting levels.  Curves are guides to
  eyes.  }
  \label{fig8}
\end{figure}
The  general behaviour  of the drift   velocity is in good qualitative
agreement with the analytical results of Ref. [\cite{rott02}] in
the   common  range of    fields.    Furthermore, recent   experiments
\cite{madhavi02} on  the drift velocity  in GaAs/GaAlAs SLs evidence a
saturation behavior of the  drift velocity for  electric fields in the
range $20  \div 50$  kV/cm, which  is  in  qualitative  agreement with
present  results.  In   this respect,  we notice  that  for   a proper
experimental validation  of  the theory more quantitative  experiments
are necessary.  In this context, we remark that present results on the
diffusion   coefficient do  not    evidence  the  Zener  antiresonance
predicted by analytical calculations \cite{kleinert04}.  The lack of a
sufficient resolution of the results due to the numerical uncertainty,
estimated to be within a  factor 2 at worst,  and/or the presence of a
substantial  carrier  heating at  the  highest fields,   could  be the
reasons for the missed evidence of  the antiresonance in the diffusion
coefficient.
\section{Conclusions}
We have    investigated  transport properties   of   an ideal infinite
superlattice in the  presence of interacting energy levels  associated
with the presence of  an applied electric  field.  To this purpose the
one dimensional  \sch\ equation in the  direction of the applied field
has been numerically solved and an ensemble Monte Carlo technique used
to  simulate the carrier motion in  the three-dimensional  states of a
perfect superlattice.  Theory   is applied to a GaAs/GaAlAs  structure
with a period of  10 nm at room temperature.   The main results of the
present work can be summarized as  follows.  (i) the interaction among
the levels is found to be responsible for a systematic increase of the
energy  levels with respect to  the bottom of the  quantum well.  (ii)
the same interaction is found to be responsible for the onset of a new
delocalization of the   Wannier-Stark  states at the  highest   fields
associated  with the tunneling  of  electrons between adjacent quantum
wells.  (iii) with  respect to the  non-interacting model, for  fields
above about 20 kV/cm  the average carrier energy  as well as the drift
velocity and    the  longitudinal  diffusion coefficient   evidence  a
consistent increase (up  to about a  factor of 10).  Such an  increase
leads to a significant softening of the negative differential value of
both  the drift   velocity  and  diffusivity  at  fields  where  Bloch
oscillation regime should be active.


%
\end{document}